\documentclass[aps,10pt,prd,citeautoscript,longbibliography,superscriptaddress,twocolumn]{revtex4-1}
\usepackage[T1]{fontenc}
\usepackage{graphicx}
\usepackage{amsmath,amssymb,amsfonts,bm,dsfont,units}
\usepackage{hyperref}
\hypersetup{colorlinks=true,citecolor=blue,linkcolor=blue,urlcolor=blue,pdfstartview=FitH}
\usepackage{braket}
\usepackage[normalem]{ulem} 
\usepackage[caption=false]{subfig} 

\newcommand{\norder}[1]{ {\mkern1mu\colon\mkern-4mu{#1}\colon\mkern-3mu} }

\newcommand{\eqnref}[1]{Eq.~(\ref{#1})}
\newcommand{\eqsref}[1]{Eqs.~(\ref{#1})}
\newcommand{\figref}[1]{Fig.~\ref{#1}}

\usepackage{color}
\definecolor{dkgreen}{rgb}{0,0.5,0}
\definecolor{midnightblue}{rgb}{0.39,0.58,0.93}

\definecolor{kspink}{RGB}{200,0,200}

\newcommand{\comment}[1]{}{}

\begin{document}

\title{Quantum spin liquids bootstrapped from Ising criticality in Rydberg arrays}

\author{Kevin Slagle}
\affiliation{Department of Physics and Institute for Quantum Information and Matter, California Institute of Technology, Pasadena, CA 91125, USA}
\affiliation{Walter Burke Institute for Theoretical Physics, California Institute of Technology, Pasadena, CA 91125, USA}
\author{Yue Liu}
\affiliation{Department of Physics and Institute for Quantum Information and Matter, California Institute of Technology, Pasadena, CA 91125, USA}
\author{David Aasen}
\affiliation{Microsoft Quantum, Microsoft Station Q, University of California, Santa Barbara, California 93106 USA}
\affiliation{Kavli Institute for Theoretical Physics, University of California, Santa Barbara, California 93106, USA}
\author{Hannes Pichler}
\affiliation{Department of Physics and Institute for Quantum Information and Matter, California Institute of Technology, Pasadena, CA 91125, USA}
\affiliation{Institute for Theoretical Physics, University of Innsbruck, 6020 Innsbruck, Austria}
\affiliation{Institute for Quantum Optics and Quantum Information, Austrian Academy of Sciences, 6020 Innsbruck, Austria}
\author{\mbox{Roger S. K. Mong}}
\affiliation{Department of Physics and Astronomy and Pittsburgh Quantum Institute, University of Pittsburgh, Pittsburgh, PA 15260, USA}
\author{Xie Chen}
\affiliation{Department of Physics and Institute for Quantum Information and Matter, California Institute of Technology, Pasadena, CA 91125, USA}
\affiliation{Walter Burke Institute for Theoretical Physics, California Institute of Technology, Pasadena, CA 91125, USA}
\author{Manuel Endres}
\affiliation{Department of Physics and Institute for Quantum Information and Matter, California Institute of Technology, Pasadena, CA 91125, USA}
\author{Jason Alicea}
\affiliation{Department of Physics and Institute for Quantum Information and Matter, California Institute of Technology, Pasadena, CA 91125, USA}
\affiliation{Walter Burke Institute for Theoretical Physics, California Institute of Technology, Pasadena, CA 91125, USA}

\date{\today}

\begin{abstract}

Arrays of Rydberg atoms constitute a highly tunable, strongly interacting venue for the pursuit of exotic states of matter.  We develop a new strategy for accessing a family of fractionalized phases known as quantum spin liquids in two-dimensional Rydberg arrays.  We specifically use effective field theory methods to study arrays assembled from Rydberg chains tuned to an Ising phase transition that famously hosts emergent fermions propagating within each chain.  
This highly entangled starting point allows us to naturally access spin liquids familiar from Kitaev's honeycomb model---albeit from an entirely different framework.  In particular, we argue that finite-range repulsive Rydberg interactions, which frustrate nearby symmetry-breaking orders, can enable coherent propagation of emergent fermions between the chains in which they were born.  Delocalization of emergent fermions across the full two-dimensional Rydberg array yields a gapless $\mathbb{Z}_2$ spin liquid with a single massless Dirac cone.  Here, the Rydberg occupation numbers exhibit universal power-law correlations that provide a straightforward experimental diagnostic of this phase.  We further show that explicitly breaking symmetries perturbs the gapless spin liquid into gapped, topologically ordered descendants:  Breaking lattice symmetries generates toric-code topological order, whereas introducing chirality generates non-Abelian Ising topological order.  In the toric-code phase, we analytically construct microscopic incarnations of non-Abelian defects, which can be created and transported by dynamically controlling the atom positions in the array.  Our work suggests that appropriately tuned Rydberg arrays provide a cold-atoms counterpart of solid-state `Kitaev materials' and, more generally, spotlights a new angle for pursuing experimental platforms for Abelian and non-Abelian fractionalization.

\end{abstract}

\maketitle

\section{Introduction}
\label{intro}

Rydberg atoms trapped in optical tweezer arrays have emerged as one of the premier platforms for engineered quantum many-body physics. Such systems display an appealing set of virtues including exceptional coherence, control over the locations of atoms in the array, the presence of strong dipole-dipole interactions, and access to site-resolved measurements. Together, these features open a new window into designing and probing novel quantum phenomena \cite{Browaeys2020,Morgado2021}. Rydberg arrays have indeed been established as fertile ground for realizing various quantum critical points \cite{Ebadi2021}, a rich set of broken-symmetry phases \cite{Lukin2017,Scholl2021,Ebadi2021}, symmetry-protected topological states \cite{RydbergSPT}, and 
long-range-entangled phases dubbed quantum spin liquids that host fractionalized excitations \cite{Verresen2020,RydbergSpinLiquid}.  

Among the preceding list, quantum spin liquids comprise an especially tantalizing target for Rydberg engineering.  Spin liquids were originally proposed as quantum-disordered ground states of frustrated magnetic insulators and have since been intensely pursued in a wide variety of materials \cite{Savary2016,ZhouQSLreview,KnolleQSLreview,BroholmQSLreview,ChamorroQSLreview}.  Rydberg arrays offer a complementary venue for tailored implementation of quantum spin liquid models, while also admitting new experimental knobs for characterization and manipulation.
As an inspiring illustration, Ref.~\onlinecite{Verresen2020} predicted the appearance of a spin liquid exhibiting toric-code topological order in a ruby-lattice Rydberg array;
remarkably,
measurements obtained shortly thereafter observed signatures of such a spin liquid state \cite{RydbergSpinLiquid,Giudici2022, Cheng2022}.  Certain classes of spin liquids, moreover, exhibit topologically protected degeneracies that enable fault-tolerant storage and manipulation of quantum information---providing longer-term motivation for stabilizing spin liquids in the highly tunable Rydberg platform.

We propose a new route to stabilizing a family of quantum spin liquids in two-dimensional (2D) Rydberg arrays.  Our approach bootstraps off of 
the limit in which a 2D array arises from initially decoupled chains, each tuned to a continuous Ising phase transition (which separates
a trivial symmetric phase from a $\mathbb{Z}_2$ charge density wave with Rydberg excitations on every other site).  This starting point confers several assets: 
Each chain is gapless and highly entangled, suggesting that exotic 2D phases ought to be nearby.
More precisely, the gapless degrees of freedom at Ising criticality include emergent fermions born within the bosonic Rydberg chains; our attack angle thus appears especially well-suited for capturing 2D phases for which the microscopic bosons fractionalize into deconfined fermions (among other nontrivial quasiparticles).  Finally, a prequel to the present study derived a  dictionary relating microscopic Rydberg operators to low-energy degrees of freedom at Ising criticality \cite{SlagleRydberg}---which is nontrivial given that the underlying Rydberg Hamiltonian is not exactly solvable.
This dictionary allows us to connect physics at all scales and, among numerous applications,
guides the identification of microscopic interactions that promote spin-liquid formation in our setups.

\begin{figure*}[t!]
   \centering
   \includegraphics[width=1.9\columnwidth]{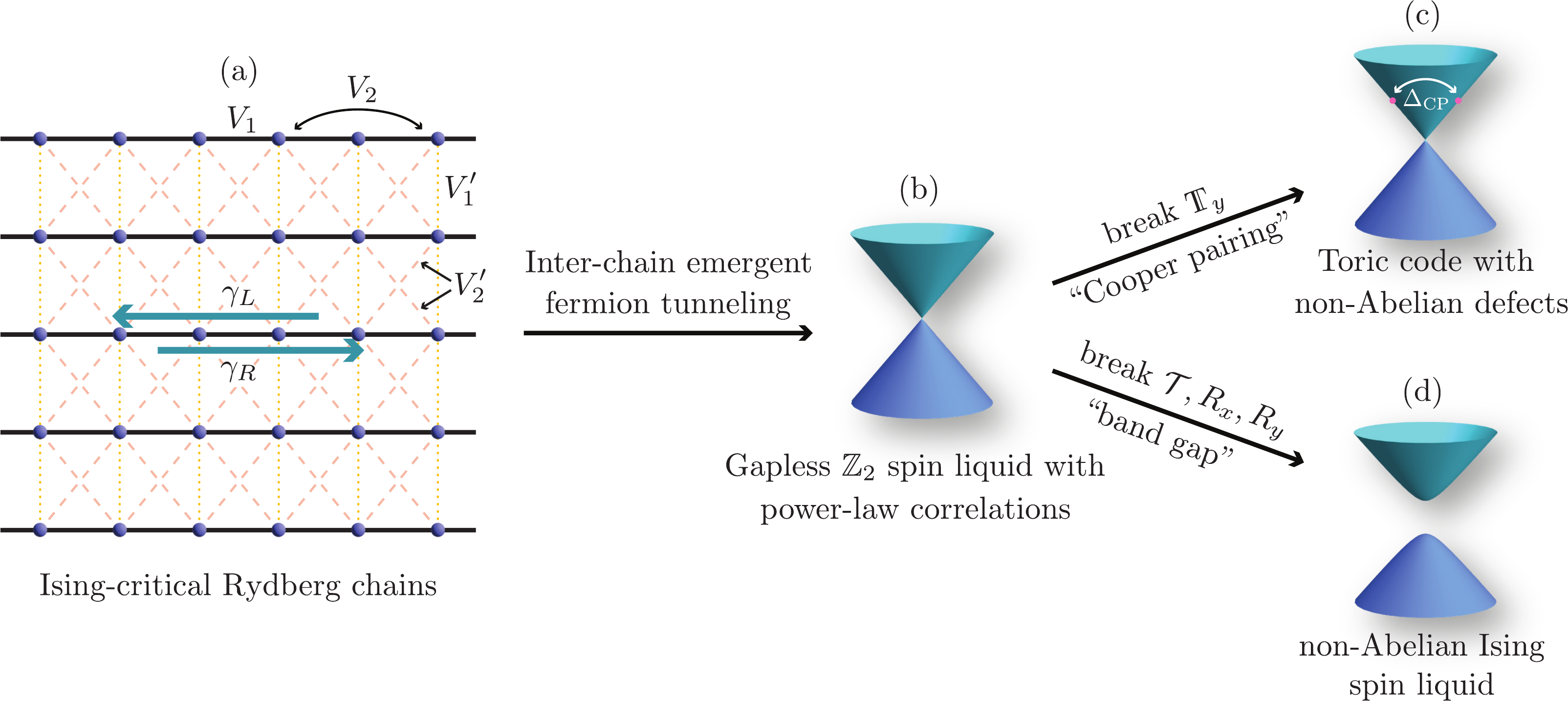}
   \caption{Executive summary.  (a) A two-dimensional Rydberg array assembled from chains tuned to Ising criticality hosts emergent Majorana fermions $\gamma_{R/L}$ propagating within each chain.  Interplay between intra-chain ($V_{1,2}$) and inter-chain ($V_{1,2}'$) repulsive Rydberg interactions promotes coherent tunneling of emergent fermions between chains, yielding (b) a gapless $\mathbb{Z}_2$ spin liquid exhibiting a single massless Dirac cone.  Explicitly breaking symmetries yields descendant gapped spin liquids: (c) Breaking inter-chain translation symmetry $\mathds{T}_y$ produces toric-code topological order, while (d) breaking time-reversal symmetry $\mathcal{T}$ and intra- and inter-chain reflection symmetries $R_{x,y}$ produces non-Abelian Ising topological order.  Either descendant can be harnessed for fault-tolerant quantum information processing.}
   \label{fig:RydbergSL}
\end{figure*}

Coupled critical Rydberg chains admit various nearby phases, including conventional symmetry-breaking orders.  Finite-range Rydberg interactions frustrate the development of symmetry breaking, however, and thus pave the way to more interesting possibilities summarized in Fig.~\ref{fig:RydbergSL}.  Specifically, we argue that the interplay between physically relevant intra- and inter-chain repulsive interactions can catalyze 
spontaneous tunneling of emergent fermions \emph{between} the critical chains comprising the array.  We motivate this outcome via an intuitive mean-field argument that holds with arbitrarily many chains, and validate the argument explicitly using bosonization in the two-chain limit. Numerical confirmation in specific microscopic models, however, is left for future work.  
In the 2D case, spontaneous inter-chain fermion tunneling 
generates a quantum spin liquid that supports gapped $\mathbb{Z}_2$ flux excitations together with \emph{gapless} fermionic excitations characterized by a single massless Dirac cone [Fig.~\ref{fig:RydbergSL}(a,b)].  (The qualitatively different approach used in Refs.~\onlinecite{Verresen2020,RydbergSpinLiquid}, by contrast, accesses only gapped topological phases.) Gaplessness is a major boon regarding experimental verification: The Rydberg occupation numbers in the 2D array in turn exhibit universal power-law correlations that constitute a clear-cut, readily measured spin-liquid signature.

Explicitly breaking symmetries 
converts the gapless spin liquid into gapped descendant topological orders suitable for fault-tolerant quantum computation \cite{Kitaev:2003,TQCreview}.
Breaking translation symmetry along the inter-chain direction---e.g., by spatially varying the detuning or modulating the atom positions---gaps out the massless Dirac cone and produces toric-code \cite{Kitaev:2003} topological order [Fig.~\ref{fig:RydbergSL}(c)].  The toric code is an Abelian topological phase, yet supports extrinsic \emph{non-Abelian} defects (originally identified with lattice dislocations in the Wen plaquette model of toric code \cite{Bombin2010}).  We use our dictionary linking microscopic operators and low-energy fields to analytically establish that non-Abelian defects in our Rydberg setups can arise from  dislocations in either the detuning pattern or the interchain separation.  The latter realization highlights a particularly enticing pathway to creation and manipulation of non-Abelian defects, given the flexibility of real-time control over the Rydberg array geometry.
Breaking time-reversal and reflection symmetries instead destabilizes the gapless spin liquid into an Ising topological order, which features a chiral Majorana edge state and bona fide deconfined non-Abelian anyons [Fig.~\ref{fig:RydbergSL}(d)].  We propose realizing this phase by periodically modulating the Rydberg atom detunings in both space and time in a manner that generates chirality on average.

The gapless spin liquid and descendant topological orders that we access appear also in a wildly different setting: Kitaev's celebrated honeycomb model \cite{Kitaev2006} for spin-1/2 degrees of freedom with bond-dependent interactions designed to enable exact solvability.  
In a seminal 2009 paper, Jackeli and Khaliullin \cite{Jackeli2009} predicted that a class of spin-orbit-coupled Mott insulators emulate Kitaev's honeycomb model, up to inevitable, nominally small perturbations.   
Exploration of such `Kitaev materials' has since evolved into a vibrant experimental enterprise (for reviews see Refs.~\onlinecite{Winter2017,Trebst2017,Hermanns18,Janssen2019,TakagiQSLreview,Motome2019}).
Our work highlights Rydberg arrays as an alternative setting in which to pursue the phenomenology of Kitaev materials---despite the fact that the microscopic interactions bear no relation to Kitaev's honeycomb model. 
We speculate that exploiting coupled critical Ising chains as done here could reveal new experimental avenues towards spin liquids in magnetic insulators as well.

To set the stage, we proceed in Sec.~\ref{Single_chain} by reviewing Ising criticality in a single Rydberg chain.  Section~\ref{ladders} then examines a Rydberg ladder assembled from a pair of critical chains.  As alluded to above, the two-chain limit allows us to controllably capture spontaneous tunneling of emergent fermions between chains. 
Although inter-chain fermion tunneling does not generate new phases in the two-chain model, it does foreshadow fractionalization in the full 2D problem.
Armed with those insights, Sec.~\ref{gaplessSL} 
characterizes the gapless spin liquid triggered by spontaneous inter-chain fermion tunneling in the 2D limit, while Sec.~\ref{descendants} investigates the gapped descendant topological orders.  
We conclude with a summary and outlook in Sec.~\ref{discussion}.

\section{Review of Ising criticality in a single Rydberg chain}
\label{Single_chain}

Reference~\onlinecite{SlagleRydberg} performed a detailed microscopic characterization of Ising criticality in a single Rydberg chain.  Here we briefly review aspects of the analysis that will be essential for later sections.  Let $j$ label sites of the chain.  Each atom has two relevant states, corresponding to the ground state (occupation number $n_j = 0$) and an excited Rydberg state (occupation number $n_j = 1$).  The atoms are driven between these two levels with Rabi frequency $\Omega$ and detuning $\Delta$ from resonance.  When populating the Rydberg excited state, pairs of atoms experience strong, long-range induced dipole-dipole interactions that suppress simultaneous Rydberg excitation among neighboring atoms.  

We model the chain's dynamics with the Hamiltonian
\begin{equation}
  H = \sum_j\left[\frac{\Omega}{2}(b_j + b_j^\dagger) -\Delta n_j + V_1 n_j n_{j+1} + V_2 n_j n_{j+2}\right],
  \label{H_FSS}
\end{equation}
where $b_j$ is a hard-core boson operator and $n_j = b_j^\dagger b_j$  \cite{Fendley2004}.
The first two terms encode driving of the atoms within the rotating-wave approximation, while $V_1$ and $V_2$ characterize first- and second-neighbor dipole-dipole interactions.
Since we focus on an effective field theory description,
  we neglect longer-range interactions and
  take $V_1 \rightarrow +\infty$; this limit enforces the `Rydberg blockade' constraint $n_j n_{j+1} = 0$ that rigidly prevents nearest-neighbor atoms from simultaneously entering the Rydberg state. 
In the following, we refer to \eqnref{H_FSS} as the Rydberg chain Hamiltonian.
  In practice $V_2$ is naturally repulsive, i.e., $V_2>0$,  though in this section it will be illuminating to also consider attractive $V_2<0$.  Equation~\eqref{H_FSS} is invariant under translation $\mathds{T}_x$ ($j \rightarrow j + 1$), reflection $R_x$ ($j \rightarrow -j$), and time reversal $\mathcal{T}$ (implemented as complex conjugation) symmetries.

\begin{figure}
  \centering
  \includegraphics[width=\columnwidth]{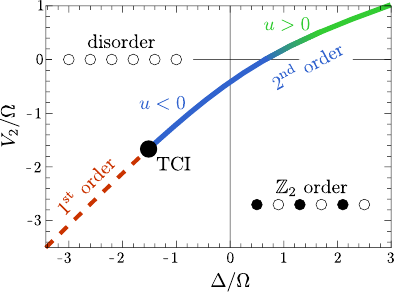}
  \caption{Phase diagram of a Rydberg chain Hamiltonian [\eqnref{H_FSS}] in the Rydberg-blockaded $V_1 \to \infty$ limit.
  }\label{fig:PhaseDiagram}
\end{figure}

Figure~\ref{fig:PhaseDiagram} reproduces the phase diagram for $H$ over a select window of $V_2/\Omega$ and $\Delta/\Omega$.   Two phases appear prominently: $(i)$ a symmetric, disordered phase smoothly connected to a trivial product state with $n_j = 0$ for all $j$, and $(ii)$ a two-fold-degenerate $\mathbb{Z}_2$ charge density wave with Rydberg excitations on every other site.  Along the phase boundary indicated by the solid line in Fig.~\ref{fig:PhaseDiagram}, the transition is continuous and falls into the Ising university class.  For sufficiently attractive $V_2<0$, however, the continuous Ising transition terminates at a tricritical Ising (TCI) point, and then becomes first order along the dashed line in Fig.~\ref{fig:PhaseDiagram}.   

The continuous phase transition line is governed by an Ising conformal field theory (CFT) with central charge $c = 1/2$.  Low-energy degrees of freedom in the CFT include a spin field $\sigma$, its dual field $\mu$, and---most importantly for this paper---right- and left-moving emergent Majorana fermions $\gamma_{R/L}$.  The spin field $\sigma$ has scaling dimension $1/8$ and corresponds to a local  $\mathbb{Z}_2$ order parameter that condenses---i.e., acquires a non-zero expectation value---in the charge-density-wave phase.  The dual field $\mu$ also has dimension $1/8$ but corresponds to a (non-local) `disorder parameter' that condenses in the trivial phase.  Fusion of $\sigma$ with $\mu$ yields emergent fermions with dimension $1/2$.  Both $\sigma$ and $\mu$ inject sign changes on the fermions via
\begin{gather}
\begin{aligned}
  \sigma(x')\gamma_{R/L}(x) &= {\rm sgn}(x'-x)\gamma_{R/L}(x) \sigma(x')\\
  \mu(x')\gamma_{R/L}(x) &= {\rm sgn}(x-x')\gamma_{R/L}(x) \mu(x'),
\end{aligned} \label{signchange}
\end{gather}
which follows from the non-local anticommutation between the spin field and its dual. We normalize the fermions such that they obey 
\begin{equation}
    \{\gamma_\alpha(x),\gamma_\beta(x')\} = \frac{1}{2}\delta_{\alpha\beta}\delta(x-x').
\end{equation}

The low-energy Hamiltonian at and near the continuous Ising transition can be compactly expressed as 
\begin{align}
  \mathcal{H} = \int_x\left[m \varepsilon + v(T + \overline{T}) + u\,T \overline{T}\right].
  \label{H_CFT}
\end{align}
Here we defined a dimension-1 fermion bilinear
\begin{equation}
    \varepsilon = i \gamma_R \gamma_L
\end{equation}
and dimension-2 kinetic energy terms
\begin{equation}
      T = -i \norder{ \gamma_R \partial_x \gamma_R } , \qquad \overline{T} = i \norder{ \gamma_L \partial_x \gamma_L } 
\end{equation}
(colons denote normal ordering).  At $m = u = 0$, Eq.~\eqref{H_CFT} reduces to the critical, fixed-point Ising CFT Hamiltonian that describes gapless emergent Majorana fermions propagating within the chain at velocity $v$.  Turning on a finite mass $m \neq 0$ gaps out the fermions and moves the chain off  criticality into either the charge-density-wave or trivial phases, depending on the sign of $m$.  In the remainder of this paper we assume $m= 0$ unless specified otherwise.  Finally, the $u$ term represents the leading two-particle interaction  that couples right- and left-moving emergent fermions.  

The $uT\bar T$ interaction, though irrelevant at weak coupling for a single chain, warrants further discussion, as it plays a central role in the multi-chain problem.  We can more clearly expose the physics of this term by rewriting
\begin{equation}
      u \int_x T \overline{T} \approx \frac{u}{\delta x^2} \int \varepsilon(x+\delta x)\varepsilon(x) + \cdots,
  \label{u_rewriting}
\end{equation}
where the ellipsis denotes fermion bilinears and a constant.  (Taylor expanding the right side up to second order in $\delta x$ yields the original form of $uT \bar T$.) From the term explicitly displayed on the right side of Eq.~\eqref{u_rewriting}, one can anticipate that a sufficiently negative $u$ drives an instability wherein $\langle \varepsilon \rangle \neq 0$.  The system then acquires acquires a gap due to spontaneous generation of a mass $m \neq 0$ whose sign, crucially, is arbitrary.  Charge-density-wave and disordered phases remain on equal footing, indicating that the continuous transition has become first order.  In sharp contrast, $u>0$ \emph{disfavors mass generation} by energetically penalizing states with $\langle\varepsilon\rangle \neq 0$.    

Reference~\onlinecite{SlagleRydberg} solidifed the connection between the Rydberg chain and the Ising critical theory in part by deriving lattice incarnations of the low-energy fields $\sigma, \mu, \gamma_{R/L}$, and $\varepsilon$.  Here we simply quote the expansion of the Rydberg number operator $n_j$ in terms of CFT fields,
\begin{equation}
    n_j \sim \langle n \rangle + c_\sigma (-1)^j \sigma + c_\varepsilon \varepsilon + \cdots,
    \label{n_expansion}
\end{equation}
where $\langle n\rangle$ is the ground-state expectation value of $n_j$, $c_\sigma$ and $c_\varepsilon$ are non-universal constants, and the ellipsis represents subleading terms with higher scaling dimension.  The $c_\sigma$ piece intuitively arises given that condensing  $\sigma$ by definition yields $\mathbb{Z}_2$ charge-density-wave order.  One can understand the $c_\varepsilon$ contribution by observing that perturbing the critical microscopic Hamiltonian with $\delta \sum_j n_j$ shifts the detuning and thus moves the chain off of criticality.  Consistency then requires that $n_j$ includes $\varepsilon$ in its expansion as quoted above.
We exploit Eq.~\eqref{n_expansion} often in later sections.

By numerically studying the spectrum for finite-size chains tuned to criticality,  Ref.~\onlinecite{SlagleRydberg} additionally quantified the dependence of CFT Hamiltonian parameters on microscopic Rydberg couplings.  Most importantly, the coefficient $u$ characterizing emergent-fermion interactions was found to vary along the continuous Ising line as indicated in  Fig.~\ref{fig:PhaseDiagram}:  Attractive $V_2<0$ yields $u<0$ couplings that eventually render the continuous Ising transition first-order, in line with the intuition provided below Eq.~\eqref{u_rewriting}.  The physically natural repulsive $V_2>0$ regime---which we focus on for the remainder of this paper---instead gives way to $u>0$ couplings that, once again, actively oppose mass generation.  This opposition is key to the onset of inter-chain fermion tunneling in multi-chain Rydberg arrays that we now turn to.

\section{Insights from the two-chain limit}
\label{ladders}

Spontaneous tunneling of emergent fermions between critical Rydberg chains is the central ingredient in our 2D spin-liquid construction.  As an instructive warm-up, this section quantifies the onset of inter-chain fermion tunneling in Rydberg ladders using mean-field arguments bolstered by a bosonization analysis.  

\subsection{Setup}
\label{Setup}

\begin{figure}[t!]
  \centering
  \includegraphics[width=.9\columnwidth]{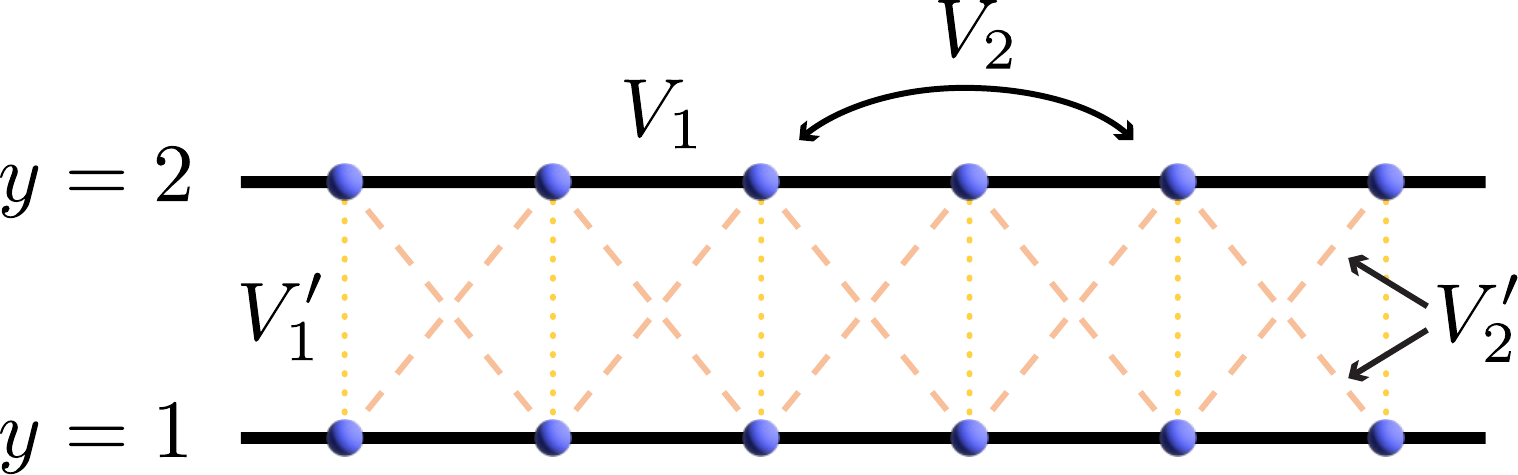}
  \caption{%
    Lattice and two-body interactions of the square Rydberg lattice Hamiltonian [\eqsref{Hsquare} and \eqref{H_2D}].
  }\label{fig:twoChain}
\end{figure}

We consider square Rydberg ladders (\figref{fig:twoChain}) built from identical chains labeled by $y = 1,2$.
We employ hard-core boson operators $b_{y,j}$ and $n_{y,j}$ for site $j$ in chain $y$ and model the system
with a microscopic Hamiltonian
\begin{equation}
  H = \sum_{y = 1,2} H_{y}^{\rm crit} + H^{\rm inter}_{1,2}.
  \label{Hsquare}
\end{equation}
In the first term, $H_y^{\rm crit}$ represents the Hamiltonian for chain $y$, given by Eq.~\eqref{H_FSS} tuned to the continuous Ising transition.  The second term captures first- and second-neighbor inter-chain density-density repulsion:
\begin{equation}
  H_{1,2}^{\rm inter} = \sum_j [V_1' n_{1,j} n_{2,j} + V_2' n_{1,j}(n_{2,j+1} + n_{2,j-1})].
\end{equation}
In addition to the symmetries enumerated previously, the Hamiltonian is invariant under a reflection $R_y$ that swaps the chains and sends $b_{1,j} \leftrightarrow b_{2,j}$, $n_{1,j} \leftrightarrow n_{2,j}$.

The expansion in Eq.~\eqref{n_expansion} allows one to straightforwardly map the inter-chain repulsion terms onto CFT fields (now denoted $\sigma_y,\mu_y, \varepsilon_y, T_y,$ etc.~for chain $y$).  This exercise  yields the following low-energy Hamiltonian for the coupled critical chains \footnote{We assume throughout that emergent fermions anticommute with one another even when they act on different chains, taking $\{\gamma_{y\alpha}(x),\gamma_{y'\beta}(x')\} = \frac{1}{2}\delta_{y,y'}\delta_{\alpha\beta}\delta(x-x')$.  Naive extension of the fermions defined for a single chain in Ref.~\onlinecite{SlagleRydberg} to the multi-chain case would yield operators that anticommute within each chain but commute between chains.   Anticommutation among fermions belonging to different chains can always be arranged by appending suitable strings to these operators.  The procedure is akin to defining Jordan-Wigner fermions in transverse-field Ising chains by adding a string that extends across the chains in a `typewriter' fashion.}:
\begin{equation}
  \mathcal{H} = \int_x \left\{\sum_{y = 1,2} \left[v\left(T_y + \bar T_y\right) + u T_y \bar T_y\right] + \kappa \sigma_1 \sigma_2 + \lambda \varepsilon_1 \varepsilon_2\right\}.
  \label{Heff_square}
\end{equation}
Technically $H_{\rm inter}$ also generates a trivial mass term $m (\varepsilon_1+\varepsilon_2)$, which we assume is cancelled by an appropriate shift of the intra-chain Hamiltonian parameters, e.g., the detuning $\Delta$.
The $\sigma_1 \sigma_2$ term, with bare coupling
\begin{equation}
  \kappa \sim V_1' -2V_2',
  \label{kappa_bare}
\end{equation}
exhibits scaling dimension $1/4$ at the decoupled-chain fixed point.  
The $\varepsilon_1 \varepsilon_2$ interaction admits bare coupling
\begin{equation}
  \lambda \sim V_1' + 2V_2'
\end{equation}
and has dimension $2$ at the decoupled-chain fixed point.
We stress that nearest- and second-neighbor inter-chain repulsion terms add up in $\lambda$ but partially cancel in $\kappa$ since longer-range interactions frustrate the development of CDW order in the chains.
Below we allow $\kappa$ to have either sign but restrict $\lambda>0$ as appropriate for inter-chain repulsion.
Note also that Eq.~\eqref{Heff_square} separately conserves the fermion parity in each chain, reflecting invariance under the gauge symmetry
\begin{equation}
    \gamma_{yR/L}\rightarrow s_y \gamma_{yR/L}
    \label{gauge_sym}
\end{equation}
with $s_y\in \pm 1$ .  
More generally, locality forbids any physical, microscopic inter-chain couplings from explicitly generating inter-chain fermion tunneling terms that would violate Eq.~\eqref{gauge_sym} and the associated conservation laws.  

\subsection{Mean-field analysis}
\label{sec:MF}

Nearby phases driven by inter-chain coupling can be accessed via a mean-field analysis of Eq.~\eqref{Heff_square}.  
When $\kappa$ dominates, the ladder develops CDW order in both chains.  The order is characterized by $\langle \sigma_1\rangle = -{\rm sgn}(\kappa)\langle \sigma_2\rangle \neq 0$, implying that the CDW's in the two chains are out of phase for $\kappa>0$ but in phase for $\kappa<0$.  

When $\lambda$ dominates, the system's fate depends on the intra-chain $u T_y \bar T_y$  interactions. As discussed in Sec.~\ref{Single_chain} and Ref.~\onlinecite{SlagleRydberg}, $uT_y \bar T_y$ promotes mass generation when $u<0$ but opposes mass generation when $u>0$.  
Suppose first that $u < 0$. Here, `large' $\lambda$ naturally drives an instability characterized by $\langle \varepsilon_1\rangle = -\langle \varepsilon_2\rangle \neq 0$, whereupon the four-fermion inter-chain interaction can be decoupled as
\begin{equation}
  \lambda \varepsilon_1 \varepsilon_2 \rightarrow m_{\rm spont}(\varepsilon_2 - \varepsilon_1)
\end{equation}
with $m_{\rm spont} = \lambda \langle \varepsilon_1 \rangle$.
Inter-chain repulsion thus spontaneously generates opposite-sign masses for the two chains---yielding a gapped phase with CDW order predominantly in one chain, which leaks CDW correlations into the other chain (due to the $\kappa$ coupling).

Now consider the $u>0$ regime that arises in the presence of physically relevant repulsive $V_2$; recall Fig.~\ref{fig:PhaseDiagram}.  
Here, suppression of mass generation by $uT_y \bar T_y$ suggests that sufficiently large $u$ qualitatively alters the instability driven by $\lambda$.  
Rewriting
\begin{align}
  \lambda \varepsilon_1 \varepsilon_2 &= -\lambda(i \gamma_{1R}\gamma_{2L})(i \gamma_{2R}\gamma_{1L}),
  \label{lambda_rewriting}
\end{align}
we see that instead condensing 
\begin{equation}
  \langle i \gamma_{1R}\gamma_{2L} \rangle = \langle i \gamma_{2R}\gamma_{1L} \rangle \neq 0
  \label{fermion_tunneling}
\end{equation}
also lowers the energy from the $\lambda$ term, but \emph{without incurring the energy penalty from $u$}.  
Condensation of this type allows us to alternatively decouple the four-fermion inter-chain interaction as
\begin{align}
  \lambda \varepsilon_1 \varepsilon_2 &\rightarrow i t_{12}(\gamma_{2R}\gamma_{1L} + \gamma_{1R}\gamma_{2L}),
  \label{decoupling}
\end{align}
yielding a state for which fermions \emph{spontaneously} tunnel between chains with amplitude $t_{12} = -\lambda \langle i \gamma_{1R}\gamma_{2L} \rangle = -\lambda \langle i \gamma_{2R}\gamma_{1L}\rangle$.   
Diagonalizing the intra-chain fermion kinetic energy supplemented by $t_{12}$ hopping reveals that the resulting state is also gapped. 

For the two-chain system considered here, the gapped state with spontaneous inter-chain fermion tunneling smoothly connects to the trivial disordered phase obtained by simply moving each chain off of criticality in the decoupled-chain limit.  In the latter regime, the trivial state exhibits condensed disordered parameters, $\langle \mu_{1,2}\rangle \neq 0$.  Turning on inter-chain coupling generically yields $\langle \sigma_1 \sigma_2\rangle \neq 0$ as well, even in the disordered phase; cf.~the $\kappa$ term in Eq.~\eqref{Heff_square}.  Consequently, $\langle i \gamma_{1R}\gamma_{2L}\rangle$---which is built from products of $\sigma_1 \sigma_2$ with disorder operators---generically condenses in the disordered phase, showing that the two states are indeed smoothly connected.  It is instructive to consider the situation in which the ladder additionally hosts a pair of microscopic $\mathbb{Z}_2$ symmetries, one that sends $\sigma_1 \rightarrow - \sigma_1, \sigma_2 \rightarrow \sigma_2$ and another that sends $\sigma_1 \rightarrow  \sigma_1, \sigma_2 \rightarrow -\sigma_2$.  These symmetries would disallow the $\kappa$ term in Eq.~\eqref{Heff_square}, implying that $\langle \sigma_1\sigma_2\rangle = 0$ in the trivial disordered phase.  Spontaneous inter-chain fermion tunneling would then trigger a cluster-state-like topological phase \cite{Son2011,Verresen2017} protected by the pair of microscopic $\mathbb{Z}_2$ symmetries.  The Rydberg chain, however, effectively realizes only the product of these $\mathbb{Z}_2$'s (implemented by single-site translation of both chains simultaneously); hence the symmetry-protected topological order is absent.

\subsection{Bosonization analysis}
\label{sec:bosonization}

An appealing feature of the two-chain ladder is that we can employ bosonization to validate the preceding analysis.  As a first step, we assemble complex right- and left-moving fermions $\psi_{R/L}$ from Majorana fermions originating in each chain and then bosonize via
\begin{align}
  \psi_R &= \gamma_{1R} + i \gamma_{2R} \sim e^{i(\varphi + \theta)} \label{psiR} \\
  \psi_L &= \gamma_{1L} + i \gamma_{2L} \sim e^{i(\varphi - \theta)}.\label{psiL}
\end{align}
Above we introduced bosonic fields $\varphi$ and $\theta$ that obey the nontrivial commutator 
\begin{equation}
    [\varphi(x),\theta(x')] = i\pi \Theta(x-x').
\end{equation}
In bosonized language, the gauge symmetry in Eq.~\eqref{gauge_sym} implies that local operators must be invariant under 
\begin{align}
    \varphi \rightarrow \varphi + \pi, \theta \rightarrow \theta
    ~~{\text{and}}~~
    \varphi \rightarrow -\varphi, \theta \rightarrow -\theta.
    \label{gauge_sym_bosonized}
\end{align}
We will also make use of time-reversal symmetry, which can be implemented as $\gamma_{yR} \leftrightarrow \gamma_{yL}$ and thus sends
\begin{equation}
    \varphi \rightarrow \varphi,  \theta \rightarrow - \theta.
    \label{T_bosonized}
\end{equation}

Next we establish a bosonization dictionary for local operators.  Majorana kinetic-energy terms in Eq.~\eqref{Heff_square} bosonize in the standard way to
\begin{align}
   \sum_{y = 1,2} v(T_y + \bar T_y) &= -i v (\psi_R^\dagger \partial_x \psi_R -  \psi_L^\dagger \partial_x \psi_L)
    \nonumber \\
    &=  \frac{v}{2\pi}[(\partial_x\varphi)^2 + (\partial_x\theta)^2].
\end{align}
We can deduce the bosonized form of the intra-chain $uT_y \bar T_y$ interactions by demanding invariance under  Eqs.~\eqref{gauge_sym_bosonized} and \eqref{T_bosonized} and matching the scaling dimension in the decoupled-chain limit; these criteria yield
\begin{align}
    \sum_{y = 1,2} u T_y\bar T_y \sim u' \cos(4\varphi) + u''\cos(4\theta)
    \label{u_bosonized}
\end{align}
with $u',u'' \propto u$.  The $\kappa$ inter-chain coupling similarly bosonizes to
\begin{equation}
    \kappa \sigma_1 \sigma_2 \sim \kappa \cos\theta.
    \label{kappa_bosonized}
\end{equation}
Although we tune explicit mass terms to zero here, it will prove useful to see how they bosonize as well.  We find it useful to directly employ Eqs.~\eqref{psiR} and \eqref{psiL} to obtain
\begin{align}
    m_s(\varepsilon_1 + \varepsilon_2) &= \frac{i}{2}m_s(\psi_R^\dagger \psi_L - \psi_L^\dagger \psi_R) \sim m_s \cos(2\theta)
    \\
    m_a(\varepsilon_1 - \varepsilon_2) &= \frac{i}{2}m_a(\psi_R^\dagger \psi_L^\dagger - \psi_L \psi_R) \sim m_a \cos(2\varphi)
    \label{m_a}
\end{align}
for mass terms that are symmetric and antisymmetric, respectively, under inter-chain reflection.
Finally, the $\lambda$ interaction may be re-expressed in terms of complex fermions as
\begin{align}
    \lambda \varepsilon_1 \varepsilon_2 = -\frac{\lambda}{4}[(\psi_R^\dagger \psi_R)(\psi_L^\dagger \psi_L) + \cdots],
    \label{lambda_intermediate}
\end{align}
where the ellipsis represents subleading terms (with extra derivatives) that do not separately conserve the number of right- and/or left-moving complex fermions.
The form on the right side indicates that repulsive inter-chain Rydberg coupling $\lambda>0$ generates \emph{attractive} density-density interactions among the $\psi_R$ and $\psi_L$ fermions.  Using the conventional expressions for the densities, $\psi_R^\dagger \psi_R \sim \partial_x(\theta + \varphi)/(2\pi)$ and $\psi_L^\dagger \psi_L \sim \partial_x(\theta- \varphi)/(2\pi)$, we obtain the bosonized form of the $\lambda$ term:
\begin{align}
    \lambda \varepsilon_1 \varepsilon_2 &\sim -\frac{\lambda}{4(2\pi)^2}[(\partial_x\theta)^2 - (\partial_x \varphi)^2] 
    \nonumber \\
    &+ \lambda' \cos(4\varphi)+ \lambda'' \cos(4\theta) + \cdots.
    \label{lambda_bosonized}
\end{align}
The last line was added to emphasize that the subleading pieces in the ellipsis of Eq.~\eqref{lambda_intermediate} can also generate terms of the form in  Eq.~\eqref{u_bosonized}.

Synthesizing the results above enables bosonization of the low-energy Rydberg ladder Hamiltonian in Eq.~\eqref{Heff_square}, yielding
\begin{align}
    \mathcal{H} &= \int_x \bigg\{ \frac{\tilde v}{2\pi}[g(\partial_x\varphi)^2 + g^{-1}(\partial_x \theta)^2]   \nonumber \\
    &+ \kappa \cos\theta+ \tilde u \cos(4\varphi) + \tilde u' \cos(4\theta)\bigg\}.
    \label{H_bosonized}
\end{align}
In the first line, $\tilde v$ is a renormalized velocity and $g$ is the Luttinger parameter that encodes density-density interactions between complex fermions originating from the inter-chain $\lambda$ interaction ($g = 1$ corresponds to the non-interacting limit).  We will focus on the $g>1$ regime appropriate for attractive interactions generated by $\lambda>0$.  In the second line, $\tilde u$ and $\tilde u'$ reflect an interplay between cosines originating from Eqs.~\eqref{u_bosonized} and \eqref{lambda_bosonized}, though we once again stress that these coefficients can be controlled by moving along the Ising transition line in Fig.~\ref{fig:PhaseDiagram}.  

Crucially, the scaling dimensions of the cosines in Eq.~\eqref{H_bosonized} vary continuously with $g$ away from their decoupled-chain values:
\begin{equation}
    [\kappa] = \frac{g}{4},~~[\tilde u] = \frac{4}{g},~~[\tilde u'] = 4g.
\end{equation}
The $\tilde u'$ term is always irrelevant in the regime of interest and will henceforth be neglected.
More interestingly, the initially strongly relevant $\kappa$ term becomes \emph{less} relevant upon increasing $g>1$ by coupling the chains; conversely, the initially irrelevant $\tilde u$ becomes relevant for $g > 2$, and beyond $g>4$ becomes even more relevant than $\kappa$.  Recall also that further-neighbor microscopic inter-chain repulsion introduces frustration that suppresses the bare value of $\kappa$ [Eq.~\eqref{kappa_bare}]---allowing $\tilde u$ to dominate even for $g<4$ if it flows to strong coupling first.  

As an aside, $\lambda$ faces less fierce competition in alternative ladder setups featuring intrinsic geometric frustration.  A zig-zag ladder, for instance, hosts a glide symmetry---corresponding to inter-chain reflection followed by translation by half a lattice spacing---that precludes the $\kappa \sigma_1\sigma_2$ term on symmetry grounds. Instead, the leading symmetry-allowed coupling between order parameters takes the form $\sigma_1 \partial_x \sigma_2$, which has a larger scaling dimension of $1+1/4$ in the decoupled-chain limit.  Multi-leg zig-zag ladders can still, however, host unfrustrated further-neighbor order-parameter couplings without derivatives \cite{Starykh2007}; thus as a worst-case scenario we only examine square ladders here.

Consider the regime in which $\kappa$ is relevant and dominates over $\tilde u$.  Here $\cos\theta$ pins $\theta$ to either $0$ or $\pi$ (depending on the sign of $\kappa$), thus fully gapping out the chains.  The mapping in Eq.~\eqref{kappa_bosonized} implies that the resulting gapped phase exhibits CDW order with $\langle \sigma_{1,2}\rangle \neq 0$.  Suppose next that $\tilde u$ is relevant and dominates over $\kappa$.  Interestingly, the system's fate now depends sensitively on the sign of $\tilde u$.  For $\tilde u < 0$, $\cos(4\varphi)$ gaps the ladder by pinning $\varphi$ to $0$ mod $\pi/2$---spontaneously generating opposite-sign masses for the two chains in accordance with the mapping in Eq.~\eqref{m_a}.  The same ordered state was argued in Sec.~\ref{sec:MF} to arise from the interplay between $\lambda$ and $u T_y \bar T_y$ interactions in the $u<0$ regime.
For $\tilde u > 0$, $\cos(4\varphi)$ gaps the ladder by instead pinning $\varphi$ to $\pi/4$ mod $\pi/2$.  Curiously, none of the local bosonized terms in Eqs.~\eqref{kappa_bosonized} through \eqref{m_a} take on non-zero expectation values upon pinning $\varphi$ in this manner.  So how should one characterize this gapped phase?  To answer this question, we rewrite the inter-chain tunneling operators from the right side of Eq.~\eqref{decoupling} in terms of complex fermions and then bosonize using Eqs.~\eqref{psiR} and \eqref{psiL}:
\begin{equation}
    i(\gamma_{2R}\gamma_{1L} + \gamma_{1R}\gamma_{2L}) = \frac{1}{2}(\psi_R \psi_L + H.c.) \sim \sin(2\varphi),
\end{equation}
where $H.c.$ denotes the Hermitian conjugate of the terms to the left.
The bosonized form above acquires a non-zero expectation value when $\varphi$ pins to $\pi/4$ mod $\pi/2$.  Consequently, this phase exhibits spontaneous inter-chain fermion tunneling, as we argued arises from an interplay between $\lambda$ and $u T_y \bar T_y$ in the physically accessible $u>0$ regime.  Note that once $\varphi$ becomes pinned, $\cos\theta$ exhibits exponentially decaying correlations and in this sense is no longer a low-energy operator.

In summary, our bosonization analysis fully corroborates the logic presented in Sec.~\ref{sec:MF}.  The key takeaway is that superficially benign, irrelevant interactions in the Ising CFT can become relevant upon coupling Rydberg chains---producing an instability in which emergent fermions are liberated from the critical chains in which they were born.  In the next section we will extend this principle to 2D Rydberg arrays to capture gapless spin liquids and their gapped topologically ordered descendants.

\section{Spin liquids in 2D Rydberg arrays}
\label{2Darrays}

\subsection{Gapless $\mathds{Z}_2$ spin liquid}
\label{gaplessSL}

We now make the leap to 2D by studying an array of $N_y \gg 1$ identical Rydberg chains (with periodic boundary conditions unless specified otherwise) governed by a straightforward generalization of the ladder Hamiltonian in Eq.~\eqref{Hsquare}:
\begin{equation}
  H_{\rm 2D} = \sum_y(H_{y}^{\rm crit} + H_{y,y+1}^{\rm inter}). \label{H_2D}
\end{equation}
As before, $H^{\rm crit}_y$ describes the intra-chain dynamics for chain $y$ and corresponds to Eq.~\eqref{H_FSS} tuned to Ising criticality, while 
\begin{equation}
  H_{y,y+1}^{\rm inter} = \sum_j [V_1' n_{y,j} n_{y+1,j} + V_2' n_{y,j}(n_{y+1,j+1} + n_{y+1,j-1})]
  \label{H_inter2D}
\end{equation}
encodes first- and second-neighbor density-density repulsion between chains $y$ and $y+1$.  For an illustration see \figref{fig:RydbergSL}(a).  Along with the symmetries present for the ladder, the full 2D  Hamiltonian preserves translations $\mathds{T}_y$ ($y \rightarrow y + 1$) along the inter-chain direction.

Rewriting $H_{\rm 2D}$ in terms of fields emerging at Ising criticality yields the low-energy Hamiltonian
\begin{align}
  \mathcal{H}_{\rm 2D} &= \int_x \sum_{y} \big[v\left(T_y + \bar T_y\right) + u T_y \bar T_y 
  \nonumber \\
  &~~~~~~~~~~~~+ \kappa \sigma_y \sigma_{y+1} + \lambda \varepsilon_y \varepsilon_{y+1}\big],
  \label{Heff_2D}
\end{align}
which similarly generalizes the continuum ladder Hamiltonian in Eq.~\eqref{Heff_square}.  In the two-chain system, we established in Sec.~\ref{ladders} that the interplay between the $\lambda$ coupling and $u T_y\bar T_y$ interactions spontaneously generated coherent tunneling of emergent fermions between chains; recall the mean-field decoupling in Eqs.~\eqref{lambda_rewriting} through \eqref{decoupling} and the supporting bosonization analysis from Sec.~\ref{sec:bosonization}.  Emboldened by those results, here we will postulate that precisely the same $\lambda$ and $u$ interplay catalyzes inter-chain fermion tunneling in the 2D Rydberg array.  That is, we work in a regime in which the decoupling 
\begin{equation}
    \lambda \varepsilon_y \varepsilon_{y+1} \rightarrow i t_{y,y+1} (\gamma_{y+1R} \gamma_{yL} + \gamma_{yR} \gamma_{y+1L})
    \label{decoupling2D}
\end{equation}
is energetically favored, where $t_{y,y+1} = - \lambda \langle i \gamma_{yR} \gamma_{y+1L}\rangle = - \lambda \langle i \gamma_{y+1R} \gamma_{yL}\rangle$ is the spontaneously generated inter-chain tunneling amplitude.  

More technically, $t_{y,y+1}(x)$ defines an emergent $\mathbb{Z}_2$ gauge field.  As encountered in the ladder problem [cf.~Eq.~\eqref{gauge_sym}], locality dictates that the 2D system preserves a gauge symmetry under $\gamma_{yR/L} \rightarrow s_y \gamma_{yR/L}$ for arbitrary signs $s_y \in \pm 1$.  This gauge symmetry in turns sends $t_{y,y+1} \rightarrow s_y s_{y+1}t_{y,y+1}$ so that the theory continues to respect locality even after employing the decoupling in Eq.~\eqref{decoupling2D}.  The crucial feature of the 2D phase with spontaneous inter-chain fermion tunneling is that $\mathbb{Z}_2$ flux excitations are gapped, promoting the fermions to bona fide deconfined 2D fractionalized excitations.  Such a phase defines a $\mathbb{Z}_2$ quantum spin liquid whose properties we will now explore.

We start by distilling Eq.~\eqref{Heff_2D} down to the free-fermion mean-field Hamiltonian
\begin{align}
    \mathcal{H}_{\rm 2D}^{\rm MF} &= \int_x \sum_y \Big[-i v\gamma_{yR}\partial_x \gamma_{yR} +iv \gamma_{yL} \partial_x \gamma_{yL} 
    \nonumber \\
    &~~~~~~~+i t(\gamma_{y+1R} \gamma_{yL} + \gamma_{yR} \gamma_{y+1L})\Big].
    \label{H2DMF}
\end{align}
The first line simply re-expresses the $T_y$ and $\bar T_y$ intra-chain kinetic energy in terms of $\gamma_{R/L}$.  To obtain the second line, we replaced  the $u$ and $\lambda$ terms with the inter-chain fermion tunneling that they conspire to generate.  We assumed a \emph{uniform} (in $x$ and $y$) tunneling amplitude $t_{y,y+1} = t$, which is equivalent to focusing on the sector with no flux excitations and choosing a convenient gauge.  Finally, we dropped the $\kappa$ term but will comment on its effects below. 

Next we pass to momentum space using the Fourier transform convention
\begin{equation}
    \gamma_{yR/L}(x) = \frac{1}{\sqrt{2N_y}}\sum_{k_y \in (-\pi,\pi]}\int_{k_x} e^{i {\bf k}\cdot {\bf r}} \gamma_{R/L}({\bf k}).
    \label{FT}
\end{equation}
Here and below ${\bf r} = (x,y)$ is a 2D position vector.
In terms of a two-component spinor $\Gamma^\dagger({\bf k}) = [\gamma_R^\dagger({\bf k})  ~\gamma_L^\dagger({\bf k})]$ with associated Pauli matrices $\tau^{x,y,z}$, the mean-field Hamiltonian can then be expressed as
\begin{equation}
    \mathcal{H}_{\rm 2D}^{\rm MF} = \sum_{k_y \in [0,\pi]}\int_{k_x} \Gamma^\dagger({\bf k})(v k_x \tau^z - t\cos k_y \tau^y)\Gamma({\bf k}).
    \label{H2DMFk}
\end{equation}
Notice that we used the Majorana property $\gamma_{R/L}(-{\bf k}) = \gamma_{R/L}^\dagger({\bf k})$ to write the Hamiltonian in terms of a sum over non-negative momenta $k_y$.  The band energies immediately follow as 
\begin{equation}
    E_\pm({\bf k}) = \pm  \sqrt{(v k_x)^2 + (t \cos k_y)^2}.
\end{equation}
In the ground state, all negative-energy $E_-({\bf k})$ states are populated while the positive-energy $E_+({\bf k})$ states are vacant.  These two bands touch at a massless Dirac point located at momentum ${\bf Q} = (0, \pi/2)$ (in the expressions that follow we do not use this explicit form of ${\bf Q}$ since the Dirac point momentum can shift in response to symmetry-preserving microscopic perturbations; see below). Consequently, the $\mathbb{Z}_2$ spin liquid hosts gapless emergent fermion excitations characterized by a \emph{single} Dirac cone as sketched in Fig.~\ref{fig:RydbergSL}(b). 

Upon  defining a low-energy spinor $\psi^\dagger({\bf q}) = [\psi_R^\dagger({\bf q})~\psi_L^\dagger({\bf q})] =  \sqrt{N_y}\Gamma^\dagger({\bf Q} + {\bf q})$, focusing on small ${\bf q}$, and Fourier transforming back to real space, Eq.~\eqref{H2DMFk} reduces to a standard (anisotropic) Dirac equation,
\begin{equation}
  \mathcal{H}_{\rm 2D}^{\rm MF} \approx \int_{\bf r} \psi^\dagger(-i v_x \tau^z \partial_x -i v_y \tau^y \partial_y)\psi.
  \label{Dirac_equation}
\end{equation}
The velocities are given by $v_x = v$ and $v_y = t$ in our units where the inter-chain spacing is set to 1.  Similarly filtering out high-energy modes from the right side of Eq.~\eqref{FT} yields the expansion
\begin{align}
    \gamma_{yR}(x) &\sim \frac{1}{\sqrt{2}}\left[e^{i{\bf Q}\cdot{{\bf r}}}\psi_R({\bf r}) + e^{-i{\bf Q}\cdot{{\bf r}}}\psi^\dagger_R({\bf r})\right]
    \label{Rexpansion}
    \\
    \gamma_{yL}(x) &\sim \frac{1}{\sqrt{2}}\left[e^{i{\bf Q}\cdot{{\bf r}}}\psi_L({\bf r}) + e^{-i{\bf Q}\cdot{{\bf r}}}\psi^\dagger_L({\bf r})\right]
    \label{Lexpansion}
\end{align}
of our original $\gamma_{y R/L}(x)$ fermions onto Dirac fields.

Gaplessness of emergent fermions in the 2D spin liquid translates into power-law correlations for physical observables in the Rydberg array.  Consider in particular the microscopic Rydberg operators $n_{y,j}$, which are most readily measurable in experiment.  Equation~\eqref{n_expansion} expresses these operators in terms of CFT fields relevant at Ising criticality.  Using Eqs.~\eqref{Rexpansion} and \eqref{Lexpansion} to further expand the CFT field $\varepsilon_y = i \gamma_{Ry}\gamma_{Ly}$ in terms of low-energy Dirac fields, we obtain
\begin{align}
    n_{y,j} \sim  \langle n\rangle - \frac{c_\varepsilon}{2}\left[\psi^\dagger \tau^y \psi
    +\left(ie^{2i{\bf Q}\cdot{{\bf r}}}\psi_L\psi_R + H.c.\right)\right].
    \label{n_expansion2D}
\end{align}
Equation~\eqref{n_expansion2D} yields the equal-time correlation function
\begin{equation}
\begin{aligned}
    &\langle n_{y,j} n_{y',j'}\rangle - \langle n_{y,j}\rangle \langle n_{y',j'}\rangle 
    \\
    &\quad \sim \frac{C_0 + C_{2{\bf Q}}\cos[2{\bf Q}\cdot (\mathbf{r} - \mathbf{r'}) + \vartheta]}{|V^{-1} \cdot(\mathbf{r} - \mathbf{r'})|^4}
\end{aligned} \label{nn_correlator}
\end{equation}
with $C_0<0, C_{2{\bf Q}}$, and $\vartheta$ non-universal constants and
\begin{equation}
  V = \begin{pmatrix} v_x & 0 \\ 0 & v_y \end{pmatrix}.
\end{equation}  

\begin{figure}[t!]
  \centering
  \includegraphics[width=.95\columnwidth]{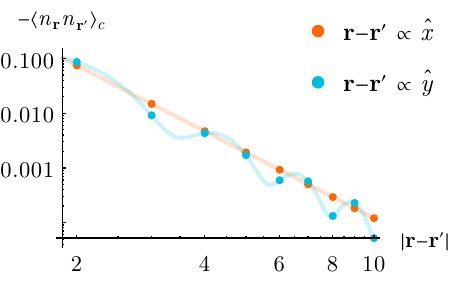}
  \caption{%
    Power-law behavior of the connected correlation function $\langle n_{\bf r} n_{\bf r'}\rangle_c$ predicted for the gapless $\mathbb{Z}_2$ spin liquid [\eqnref{nn_correlator}]. 
   Data were obtained with non-universal example parameters $\vartheta=0$, $v_x=v_y=1$, $C_0 = -1$, $C_{2{\bf Q}} = -0.5$, and ${\bf Q} = (0, 1.4)$ (though the vanishing of $Q_x$ is generic provided $R_x$ and $R_y$ reflection symmetries are preserved).
   The orange curve corresponds to separations ${\bf r-r'}$ along a chain, while the blue curve corresponds to separations perpendicular to the chains; additional oscillations in the latter reflect power-law correlations at momentum $2{\bf Q}$.
  }\label{fig:powerLaw}
\end{figure}

Equation~\eqref{nn_correlator}
reveals power-law correlations---displayed in Fig.~\ref{fig:powerLaw}---at both zero momentum and $2{\bf Q}$ with the same universal exponent.  This feature constitutes a `smoking-gun' signature of the gapless $\mathbb{Z}_2$ spin liquid.
Indeed, the only other ways to obtain power-law correlations in a bosonic model are via either fine-tuning (e.g., to a phase transition) or continuous symmetries (which our model does not exhibit).
See Sec.~\ref{discussion} for further discussion of this important point.

The gapless spin liquid also exhibits $\mathbb{Z}_2$ flux excitations.
To understand them, note that in $\mathcal{H}_{\rm 2D}^{\rm MF}$, one can introduce a $\mathbb{Z}_2$ flux excitation localized at position $x_f$ between chains $y_f$ and $y_f+1$ by modifying  $t_{y_f,y_f+1} \rightarrow {\rm sgn}(x-x_f)t$ but retaining $t_{y,y+1} = t$ elsewhere.
Fermions encircling the flux pick up a minus sign due to the sign change in $t_{y_f,y_f + 1}$. Dimensional analysis suggests that the energy cost is 
\begin{equation}
  E_{\rm flux} = {\rm const} \times \sqrt{(v_x\Lambda_x) (v_y \Lambda_y)}
  \label{Eflux}
\end{equation} with $\Lambda_x$ and $\Lambda_y \sim 1$ momentum cutoffs associated with the intra and inter-chain directions, respectively.
Dependence on $\Lambda_{x,y}$ reflects the fact that the entire filled Dirac sea `feels' the sign change engendered by the flux; moreover, the energy cost must vanish when either $v_x$ or $v_y$ vanishes, as captured by this expression.  For additional support we can appeal to a lattice system that yields the same low-energy physics---e.g., the Kitaev honeycomb model.  Equation~\eqref{Eflux} is consistent with the well-established energy gap in the latter problem.  

The CFT field $\sigma_y(x)$ enacts a sign change for the fermions in chain $y$ [recall Eq.~\ref{signchange}] and thus creates a \emph{pair} of $\mathbb{Z}_2$ fluxes, one between chains $y, y+1$ and another between chains $y-1,y$.  Accordingly, the once mighty $\sigma_y(x)$ field---highly relevant in the decoupled-chain limit---has been demoted to a gapped operator.  This conclusion justifies our neglect of $\sigma_y$ in the low-energy expansion in Eq.~\eqref{n_expansion2D}, as well as our neglect of the $\kappa$ term in Eq.~\eqref{Heff_2D}.  Reviving $\kappa$ mixes nontrivial flux configurations into the ground state, similar to the inclusion of perturbations that spoil exact solvability of the Kitaev honeycomb model.  Such processes complicate the analysis of the spin liquid but do not alter its universal properties, provided $\kappa$ is sufficiently weak.

\begin{table*}
\centering
\begin{tabular}{|c | c | c | c | c | c | c |} 
 \hline
  & $\mathds{T}_x$ & $\mathds{T}_y$ & $R_x$ & $R_y$ & $\mathcal{T}$ & physical meaning \\ 
 \hline
 $\psi \rightarrow $ & $\psi$ & $e^{i Q} \psi$ & $i \tau^y \psi$ & $[\psi^\dagger]^T$ & $\tau^x[\psi^\dagger]^T$ & low-energy Dirac spinor \\ 
 $\psi^\dagger \psi \rightarrow$ & $\psi^\dagger \psi$ & $\psi^\dagger \psi$ & $\psi^\dagger \psi$ & $-\psi^\dagger \psi$  & $-\psi^\dagger \psi$ & chemical potential \\
 $\psi^\dagger\tau^x \psi \rightarrow$ & $\psi^\dagger\tau^x \psi$ & $\psi^\dagger\tau^x \psi$ & $-\psi^\dagger\tau^x \psi$ & $-\psi^\dagger\tau^x \psi$ & $-\psi^\dagger\tau^x \psi$ & Dirac mass, generates non-Abelian spin liquid
 \\
 $\psi^\dagger\tau^y \psi \rightarrow$ & $\psi^\dagger\tau^y \psi$ & $\psi^\dagger\tau^y \psi$ & $\psi^\dagger\tau^y \psi$ & $\psi^\dagger\tau^y \psi$ & $\psi^\dagger\tau^y \psi$ & shifts Dirac point momentum ${\bf Q}$ along ${\bf \hat {y}}$
 \\
 $\psi^\dagger\tau^z \psi \rightarrow$ & $\psi^\dagger\tau^z \psi$ & $\psi^\dagger\tau^z \psi$ & $-\psi^\dagger\tau^z \psi$ & $-\psi^\dagger\tau^z \psi$ & $\psi^\dagger\tau^z \psi$ & shifts Dirac point momentum ${\bf Q}$ along ${\bf \hat {x}}$
 \\
 $\psi_L \psi_R \rightarrow$ & $\psi_L \psi_R$ & $e^{2i Q} \psi_L \psi_R$ & $\psi_L \psi_R$ & $- \psi_R^\dagger \psi_L^\dagger$ & $\psi_R^\dagger \psi_L^\dagger$ & `Cooper pairing', generates toric-code spin liquid
 \\
 \hline
\end{tabular}
\caption{Symmetry transformation properties of low-energy Dirac fields $\psi$ and fermion bilinears in the quantum spin liquid phase.  For brevity we suppress the coordinates, which should also transform appropriately under spatial symmetries (e.g., $x\rightarrow -x$ under $R_x$).
}\label{SymmetriesTable}
\end{table*}

To assess the stability of the gapless spin liquid more generally, we use Eqs.~\eqref{Dirac_equation} and \eqref{n_expansion2D} to deduce a consistent embedding of microscopic symmetries on the Dirac fields.  Table~\ref{SymmetriesTable} summarizes the action on $\psi$ as well as momentum-independent fermion bilinears.  Notice that $R_y$ and $\mathcal{T}$ both act as particle-hole transformations on the Dirac fermions. These symmetries thus preclude perturbations of the form $-\mu \psi^\dagger \psi$, i.e., they pin the chemical potential precisely to the massless Dirac point.  All microscopic symmetries do, however, permit a fermion bilinear $\propto \psi^\dagger \tau^y \psi$ in the Hamiltonian.  
Physically, such a term arises upon moving each chain off of Ising criticality by adding 
\begin{align}
    \mathcal{H}_{{\bf Q}\text{-}{\rm shift}} = m\int_x \sum_y  \varepsilon_y \sim -\frac{m}{2}\int_{\bf r} \psi^\dagger \tau^y \psi.
\end{align}
While the perturbation $m \varepsilon_y$ immediately opens a gap in the decoupled-chain limit, it does \emph{not} destabilize the 2D spin liquid phase arising in the coupled-chain system (at least for sufficiently weak $m$).  Indeed, 
when added to the Dirac Hamiltonian in Eq.~\eqref{Dirac_equation}, the fermion bilinear on the right side merely shifts the massless Dirac-point momentum to ${\bf Q} = (0, \pi/2 +v_y m/2)$.  Addition of a term $\propto \psi^\dagger \tau^z\psi$ would similarly shift ${\bf Q}$ along the $x$ direction, but is disallowed by both $R_x$ and $R_y$.

The presence of gapless fermionic excitations does, nevertheless, imply the existence of nearby gapped topological orders that we explore in the next subsection.  Gapped descendants arise from perturbing the massless Dirac equation with either a Dirac mass $M\psi^\dagger \tau^x \psi$ or a `Cooper pairing' term $(\Delta_{\rm CP} \psi_L \psi_R + H.c.)$. (We use $M$ for the 2D Dirac mass to distinguish from the mass $m$ that pushes a single chain off of Ising criticality; also, the subscript `CP' is inserted to differentiate from the microscopic detuning parameter $\Delta$.)  Either gapping mechanism requires explicitly breaking microscopic symmetries enjoyed by the Rydberg platform---i.e., the gapless $\mathbb{Z}_2$ spin liquid accessed here is stable to arbitrary \emph{symmetry-preserving} perturbations. 

\subsection{Descendant gapped spin liquids}
\label{descendants}

\subsubsection{Toric code}

We first explore the effect of Cooper pairing emergent fermions, which yields a gapped descendant phase analogous to the Fu-Kane superconductor in a 3D topological insulator surface \cite{Fu2008}.  The only symmetry precluding pairing in the continuum Dirac theory is translation $\mathds{T}_y$; recall Table~\ref{SymmetriesTable}.  Moreover, Eq.~\eqref{n_expansion2D} reveals that  lattice-scale modulations of the Rydberg occupation number relate to Dirac-fermion pairing.  We can use this correspondence to identify microscopic perturbations that open a pairing gap.  As a simple example, suppose that we add
\begin{align}
    H_{\rm TC} &=  \sum_{y,j}\widetilde{\Delta}({\bf r})\cos[2{\bf Q}\cdot {\bf r}-\Phi({\bf r})]n_{y,j}
    \nonumber \\
    &\sim  \int_{\bf r} \left[\Delta_{\rm CP}({\bf r}) e^{i \Phi({\bf r})} \psi_L \psi_R + H.c.\right] 
    \label{H_TC}
\end{align}
to the 2D Rydberg Hamiltonian, where ${\bf r} = (j,y) \sim (x,y)$, $\Delta_{\rm CP} \propto i \widetilde{\Delta}$, and $\widetilde{\Delta}({\bf r}), \Phi({\bf r})$ are some slowly varying real-valued function that we are free to specify.  The first line represents a position-dependent modulation of the original detuning $\Delta$.  Equation~\eqref{n_expansion2D} yields the second line upon dropping rapidly oscillating terms.  Importantly, we assumed the generic situation wherein ${\bf Q}$ is not fine-tuned to $(0,\pi/2)$;  notice that in this case $\Phi({\bf r})$ maps to the phase of the pairing potential in the low-energy theory.  For the special case ${\bf Q} = (0,\pi/2)$, the property $e^{2i {\bf Q}\cdot {\bf r}} = e^{-2i {\bf Q}\cdot {\bf r}}$ results in a different integrand in Eq.~\eqref{H_TC} $\propto \cos[\Phi({\bf r})](i\psi_L \psi_R + H.c.)$. The function $\Phi({\bf r})$ then instead controls the amplitude of the pairing potential rather than its phase.  

\begin{figure*}[t!]
  \centering
  \subfloat[]{\includegraphics[height=.65\columnwidth]{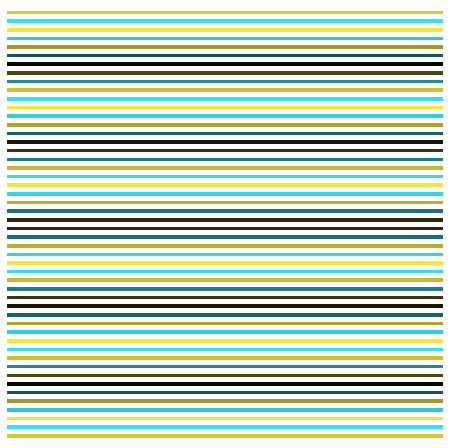}}
  \hspace{1cm}
  \subfloat[\label{fig:HTCb}]{\includegraphics[height=.65\columnwidth]{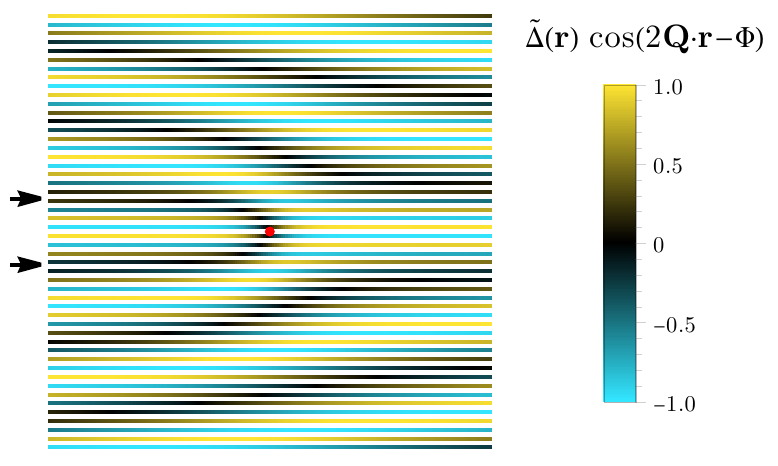}}
  \caption{%
    (a) Spatially modulated detuning pattern that generates toric-code topological order by explicitly breaking inter-chain translation symmetry $\mathds{T}_y$.  (b) Modified detuning pattern that yields a non-Abelian defect (centered at the red dot) in the toric-code phase.  The color represents the detuning shift $\widetilde{\Delta}({\bf r}) \cos[2{\bf Q}\cdot {\bf r}-\Phi({\bf r})]$ from \eqnref{H_TC} with
    ${\bf Q} = (0, 1.4)$ in both panels.  In (a), we use uniform $\widetilde{\Delta}({\bf r})$ and $\Phi({\bf r})$. In (b), we instead take $\Phi({\bf r})$ to wind by $2\pi$ around the red dot and set $\widetilde{\Delta}({\bf r}) \propto \tanh(r/\xi)$ (with $\xi^{-1} 
    \approx 1/3$) to introduce a vortex core.  The colorbar is normalized such that $\pm 1$ correspond to the maximum and minimum values of the detuning shift---which should reflect small perturbations to the microscopic Hamiltonian.
  }\label{fig:HTC}
\end{figure*}

Addition of the microscopic perturbation in Eq.~\eqref{H_TC} with spatially uniform $\widetilde{\Delta}({\bf r})$ and $\Phi({\bf r})$ breaks translation symmetry along $y$ in a manner that turns the gapless $\mathbb{Z}_2$ spin liquid into a fully gapped phase that we identify as the toric code [Fig.~\ref{fig:RydbergSL}(c)]. The toric code hosts three nontrivial quasiparticle types: $e$ particles, $m$ particles, and fermions $f = e \times m$ that emerge from fusion of $e$ and $m$.  Fusing two quasiparticles of the same type yields a local boson denoted $I$, i.e., $e\times e = m \times m = f \times f = I$.  Both $e$ and $m$ are self-bosons but mutual semions; it follows that taking a fermion around either $e$ or $m$ yields a statistical phase of $-1$.   
In our pairing-gapped spin liquid, the presence of fermions is obvious given the massive Dirac cone, while $\mathbb{Z}_2$ flux excitations can be identified with $m$ particles.  A fermion encircling a flux picks up a minus sign as remarked earlier, thus accounting for the mutual statisics between $f$ and $m$.  The toric code's $e$ particles arise from bound states of fermions and $\mathbb{Z}_2$ fluxes.  

Although $e, m,$ and $f$ exhibit only Abelian braiding statistics, the toric code is known to admit extrinsic non-Abelian defects, which are non-dynamical cousins of Ising anyons that bind Majorana zero modes.
In the anyon model, the non-Abelian defect can be viewed as a branch cut that swaps $e \leftrightarrow m$ particles \cite{Bombin2010}.
We can analytically pinpoint non-Abelian defects in our Rydberg  toric-code realization: In the generic case ${\bf Q} \neq (0,\pi/2)$, they arise upon replacing the spatially uniform $\Phi({\bf r})$ function considered above with a non-uniform profile that yields a single-strength vortex in the effective pairing potential from the second line of Eq.~\eqref{H_TC}.  [To avoid pathologies one should also replace the spatially uniform $\widetilde{\Delta}({\bf r})$ with a profile that vanishes at the vortex core.] The vortex indeed binds a Majorana zero mode, which follows immediately from the analogy with the Fu-Kane superconductor \cite{Fu2008}.  The first line of Eq.~\eqref{H_TC} indicates that the non-Abelian defect can be created by introducing a judicious spatially modulated detuning pattern.

Figure~\ref{fig:HTC} illustrates the detuning pattern (a) without and (b) with a vortex.  The color associated with each chain indicates the local detuning shift $\tilde \Delta({\bf r})\cos[2{\bf Q}\cdot {\bf r}-\Phi({\bf r})]$.  
Inspection of Fig.~\ref{fig:HTC}(b) reveals that a vortex induces a \emph{pair} of edge dislocations in the detuning profile.
That is, viewing the darker regions that have near-zero detuning shifts as `super-layers,'
  we see that the left side has two more super-layers
  (marked by arrows in \figref{fig:HTCb}) than the right side.
A single dislocation, incidentally, is not possible by tuning $\Phi({\bf r})$.
Neighboring blue/yellow-colored super-layers exhibit a relative $\pi$-shift in $\Phi({\bf r})$.
This feature can be seen most easily near a vortex (red dot),
  horizontally across which each Rydberg layer changes color from yellow to blue in \figref{fig:HTCb}.
Thus defects induced by $\Phi$ can only change the number of super-layers by an even number, ruling out single dislocations.  

\begin{figure*}[t!]
  \centering
  \subfloat[\label{fig:HTCp0}]{\includegraphics[height=.65\columnwidth]{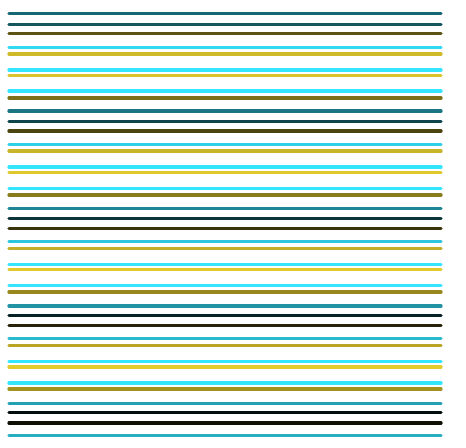}}
  \hspace{1cm}
  \subfloat[\label{fig:HTCp}]{\includegraphics[height=.65\columnwidth]{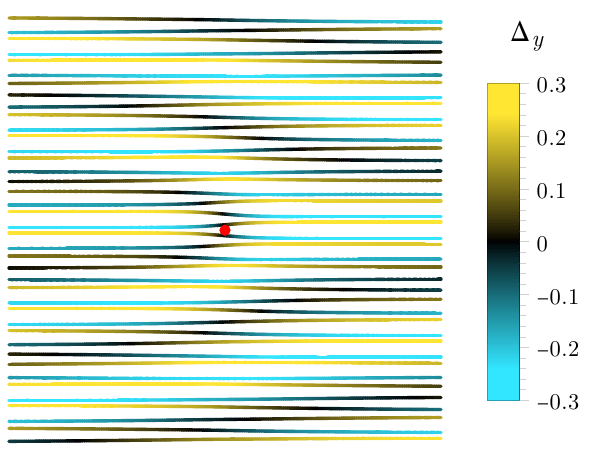}}
  \caption{%
    Variation of Fig.~\ref{fig:HTC} that accesses toric-code topological order by breaking inter-chain translation symmetry $\mathds{T}_y$ via small shifts in the Rydberg-chain positions [leading to Eq.~\eqref{H_TCp}].  Panel (a) realizes the defect-free toric code, while panel (b) hosts a non-Abelian defect centered at the red circle.  Chains are colored according to their displacement $\Delta_y$ away from their original positions (exaggerated for clarity and in units of interchain spacing).  The positions were calculated by assuming $\Delta_{y+1} - \Delta_y \propto - A({\bf r}) \cos[2 {\bf Q} \cdot {\bf r} - \Phi({\bf r})]$ with ${\bf Q} = (0,1.4)$ as in Fig.~\ref{fig:HTC}.  In (a) both $A({\bf r})$ and $\Phi({\bf r})$ are uniform, while in (b) $\Phi({\bf r})$ winds by $2\pi$ around the red circle and $A({\bf r}) = \tanh(r/\xi)$ with $\xi^{-1} \approx 1/3$.  
  }\label{fig:HTCps}
\end{figure*}

The preceding route to the toric code is by no means unique.  As a perhaps more practical variation, consider the alternative microscopic perturbation
\begin{align}
    H_{\rm TC}' &= \sum_{y,j}\cos[2{\bf Q}\cdot {\bf r}-\Phi({\bf r})][\delta V_1'({\bf r}) n_{y,j} n_{y+1,j} 
    \nonumber \\
    &+ \delta V_2'({\bf r}) n_{y,j}(n_{y+1,j+1} + n_{y+1,j-1})]
    \nonumber \\
    &\sim  \int_{\bf r} \left[\Delta_{\rm CP}({\bf r}) e^{i \Phi({\bf r})} \psi_L \psi_R + H.c.\right]
    \label{H_TCp}
\end{align}
that encodes spatial variation in the inter-chain couplings from Eq.~\eqref{H_inter2D}; here $\delta V_{1,2}'({\bf r}), \Phi({\bf r})$ are slowly varying functions. 
To arrive at the bottom line, we again assumed ${\bf Q} \neq (0,\pi/2)$, yielding a pairing term in the Dirac Hamiltonian with $\Delta_{\rm CP} \propto i(e^{2i{\bf Q}\cdot{\bf \hat{y}}} + 1)(\delta V_1' + 2\delta V_2')$ and phase controlled by $\Phi({\bf r})$.  The microscopic incarnation from Eq.~\eqref{H_TCp} benefits from a simple physical implementation:
the encoded modulations arise upon tweaking the atom positions in a way that alters the local inter-chain separation, e.g., bringing a pair of chains locally closer enhances their inter-chain repulsion whereas separating them decreases their repulsion. 
Consider the `pure' toric-code phase emerging at constant $\delta V_{1,2}'({\bf r})$ and $\Phi({\bf r})$.  In that limit the upper lines of Eq.~\eqref{H_TCp} arise from modulating the inter-chain spacing along $y$ with wavelength $\lambda = 2\pi/|2{\bf Q}|$.   Figure~\ref{fig:HTCp0} illustrates such a modulation, with the colors indicating the chain displacements $\Delta_y$ relative to their original positions.  Introducing the non-Abelian defect via non-uniform $\Phi({\bf r})$ that adds a vortex [with an accompanying profile in $\delta V'_{1,2}({\bf r})$ that vanishes at the core] yields the deformed geometry shown in Fig.~\ref{fig:HTCp}.
Similar to \figref{fig:HTCb}, the vortex results in a pair of dislocations in the super-layer pattern (darker regions in figure).

Figure~\ref{fig:HTCps} was generated by assuming small, slowly varing changes in the inter-chain spacings relative to the uniform case (although we exaggerate the magnitude of the modulation in the figure to improve readability).
The change in inter-chain interaction strength can then be approximated as linearly proportional to the local change in inter-chain spacing $\Delta_y(x)$.
That is, we take $\delta V_{1,2}'({\bf r}) \cos[2{\bf Q}\cdot {\bf r}-\Phi({\bf r})] \propto - [\Delta_{y+1}(x)-\Delta_y(x)]$ 
  with a proportionality constant arbitrarily selected for ease of visualization. 
One can then infer (modulo a trivial overall shift) the individual $\Delta_y(x)$ shifts from a given $\delta V_{1,2}'({\bf r}), \Phi({\bf r})$ profile, leading to Fig.~\ref{fig:HTCps}.

\subsubsection{Ising topological order}

Adding a Dirac mass term $M \psi^\dagger \tau^x \psi$ for the emergent fermions requires breaking $\mathcal{T}, R_x,$ and $R_y$ (see Table~\ref{SymmetriesTable}), and  
produces a descendant spin liquid with emergent fermions analogous to those of the magnetically gapped surface of a 3D topological insulator \cite{QAH_TI}.  This analogy takes us far in determining the nature of the corresponding state in the Rydberg setting: 
The interface between domains gapped by pairing and by a mass term hosts a single chiral Majorana mode \cite{FuKaneInterferometer,AkhmerovInterferometer}.  Since pairing generates a time-reversal-invariant, Abelian toric-code phase for the Rydberg array, the chiral Majorana mode can only arise as an edge state of the gapped spin liquid generated by the Dirac mass.  This property identifies the latter phase as a \emph{non-Abelian} spin liquid with Ising topological order [Fig.~\ref{fig:RydbergSL}(d)].  

\begin{figure}
   \centering
   \includegraphics[width=.7\columnwidth]{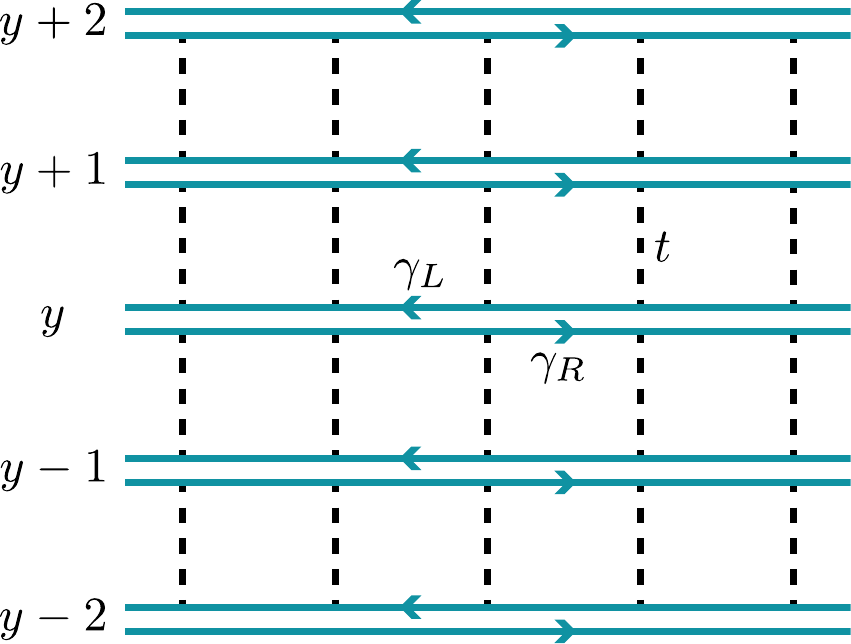}
   \caption{Pattern of interchain emergent-fermion tunnelings encoded in the second line of Eq.~\eqref{H_chiral}. The left-mover originating from chain $y$ couples to the right-mover from chain $y+1$, gapping the interior but leaving unpaired chiral Majorana edge modes at the top and bottom chains.  In the 2D Rydberg setup, the resulting phase is a non-Abelian spin liquid with Ising topological order.}
   \label{fig:CoupledChain}
\end{figure}

For a more direct treatment, we use Eqs.~\eqref{Rexpansion} and \eqref{Lexpansion} to obtain
\begin{equation}
    \delta \mathcal{H} \equiv \int_x \sum_y i\delta t(\gamma_{y+1R}\gamma_{yL} -  \gamma_{yR}\gamma_{y+1L}) \sim \int_{\bf r} M\psi^\dagger \tau^x \psi
\end{equation}
with $M \propto \delta t$.  Adding $\delta \mathcal{H}$ to the mean-field Hamiltonian $\mathcal{H}_{\rm 2D}^{\rm MF}$ from Eq.~\eqref{H2DMF} increases the amplitude for $i\gamma_{y+1R}\gamma_{yL}$ inter-chain hopping but decreases the amplitude for $i\gamma_{yR}\gamma_{y+1L}$ hopping (or vice versa depending on the sign of $\delta t$).  The illuminating special case $t = \delta t$ vanquishes $i \gamma_{yR}\gamma_{y+1L}$ tunneling altogether and yields
\begin{align}
    \mathcal{H}_{\rm 2D}^{\rm MF} + \delta \mathcal{H} &= \int_x \sum_y (-i v\gamma_{yR}\partial_x \gamma_{yR} +iv \gamma_{yL} \partial_x \gamma_{yL} 
    \nonumber \\
    &~~~~~~~+2i t\gamma_{y+1R} \gamma_{yL}).
    \label{H_chiral}
\end{align}
Figure~\ref{fig:CoupledChain} sketches the pattern of inter-chain tunnelings that remain \cite{TeoKane}, which gap the bulk by hybridizing the left-mover from chain $y$ with the right-mover from the chain above. A $\mathbb{Z}_2$ flux excitation, obtained by sending $t\rightarrow {\rm sgn}(x-x_f)t$ for one of the interchain tunnelings as before, binds a Majorana zero mode \cite{TeoKane} and correspondingly realizes the gapped non-Abelian anyon characteristic of Ising topological order.

With open boundary conditions along the inter-chain direction, the pattern of inter-chain tunnelings in Fig.~\ref{fig:CoupledChain} yields unpaired chiral  Majorana fermions: one at the top chain and one at the bottom.   These chiral modes define the hallmark gapless edge states captured previously from the low-energy Dirac Hamiltonian viewpoint.  (Away from the fine-tuned limit $t = \delta t$, the chiral edge modes simply decay exponentially into the bulk rather than localizing to a single chain.)  The microscopic operator $n_{y,j}$ evaluated near a boundary admits the low-energy expansion
\begin{equation}
    n_{y,j} \sim {\rm const} + i a_{y,j} \gamma_{\rm edge} \partial_u \gamma_{\rm edge},
    \label{edge_expansion}
\end{equation}
where $\gamma_{\rm edge}$ is the gapless chiral Majorana field, $u$ is a coordinate along the edge, and $a_{y,j}$ are non-universal constants that decay exponentially as $y,j$ move into the bulk.  One can motivate Eq.~\eqref{edge_expansion} from Eq.~\eqref{n_expansion} upon explicitly including a subleading $T+\bar T$ term allowed by symmetry in the ellipsis; in the special $t = \delta t$ limit, either $T$ or $\bar T$ yields the chiral edge state kinetic energy $\propto i\gamma_{\rm edge}\partial_u \gamma_{\rm edge}$ at the outermost chains.  Equation~\eqref{edge_expansion} implies universal power-law correlations among \emph{boundary} Rydberg occupation numbers:
\begin{align}
    \langle n_{y,j}n_{y',j'}\rangle - \langle n_{y,j}\rangle \langle n_{y',j'}\rangle \sim \frac{1}{(u-u')^4},
    \label{edge_power_laws}
\end{align}
with $u-u'$ the separation between the operators along the edge.  Bulk operators, by contrast, exhibit exponentially decaying correlations.

Breaking reflection and time reversal symmetries could be achieved at a microscopic level by adding a position-- and time-dependant perturbation to the detuning:
\begin{equation}
  \delta H = \sum_{y,j} \left[ (-1)^{y+j} \widetilde{\Delta}_1 + \widetilde{\Delta}_2(t) \right] n_{y,j}.
\end{equation}
The $\widetilde{\Delta}_1$ term breaks reflection symmetries $R_x$ and $R_y$,
  but conveniently preserves the products $R_x R_y$ and $\mathds{T}_x^{-1} \mathds{T}_y$---
  which respectively guarantee that neither a chemical potential nor a `Cooper pairing' term is generated.
We assume that the time-dependant $\widetilde{\Delta}_2(t)$ perturbation is periodic with a very short period
  so that an effective Floquet Hamiltonian \cite{GoldmanDalibard2014} description is valid.
The effective Hamiltonian will generically break time reversal symmetry as long as $\widetilde{\Delta}_2(t)$ is not time-reversal-symmetric,
  i.e., as long as $\widetilde{\Delta}_2(t+t_0) \neq \widetilde{\Delta}_2(-t+t_0)$ for all $t_0$.

\section{Discussion}
\label{discussion}

Bootstrapping off of theoretically well-understood Ising criticality in Rydberg chains, we proposed a new route to quantum spin liquids in 2D Rydberg arrays.  Our approach directly targets a \emph{gapless} $\mathbb{Z}_2$ spin liquid featuring emergent fermions characterized by a single massless Dirac cone.  This `parent' state gives way to descendant gapped spin liquids---either the toric code or non-Abelian Ising topological order---upon introducing (arbitrarily weak) symmetry-breaking perturbations.  

Our construction synthesized three ingredients: $(i)$ The ability to tune a single Rydberg chain into the continuous Ising transition separating trivial and charge-density-wave phases.  $(ii)$ Control over the sign and strength of intra-chain interactions among fermions emerging at the Ising transition, which can be achieved by modulating the detuning and Rabi frequency (Fig.~\ref{fig:PhaseDiagram}).  Such interactions are irrelevant for a single chain but can nevertheless play a decisive role in determining the fate of 2D Rydberg arrays.  $(iii)$ Microscopic density-density  interactions between neighboring critical Rydberg chains. 
We argued (and showed explicitly in the two-chain limit) that the interplay between intra- and inter-chain interactions among emergent fermions enables them to coherently tunnel between chains, producing the gapless $\mathbb{Z}_2$ spin liquid in the limit of a 2D Rydberg array. 

Although we fall short of pinpointing precisely where in parameter space the gapless spin liquid phases arises, we hope that our work will stimulate numerical efforts in this direction.  Prospects here appear quite promising given that the microscopic models underlying our field-theoretic treatments do not admit a sign problem.  For simplicity we focused primarily on square-lattice geometries in this paper, though frustrated lattices (which remain sign-problem-free) appear even more amenable to our spin-liquid construction due to additional suppression of order-parameter couplings between chains; recall the discussion of zigzag ladders in Sec.~\ref{sec:bosonization}.  

Such numerical investigations are further motivated by a fundamental question at the intersection of quantum information and quantum simulation.  For certain phases of matter (e.g., the toric code), it is known that parent sign-problem-free microscopic Hamiltonians exist.  In some cases, however, (e.g., Ising topological order), the existence of a sign problem is unavoidable \cite{ZoharSignProblem,ZoharSignProblem2}. To our knowledge, whether a gapless $\mathbb{Z}_2$ spin liquid can emerge from a sign-problem-free Hamiltonian has remained an open question.  Our work suggests that such Hamiltonians do indeed exist---which would be interesting to establish from this angle alone.  As a sanity check, deforming our candidate microscopic spin-liquid Hamiltonians to the toric code phase by breaking translation symmetry does not introduce a sign problem, whereas deforming to Ising topological order does due to breaking of time-reversal symmetry.

A very appealing feature of  the gapless $\mathbb{Z}_2$ spin liquid is its ease of experimental identification in Rydberg arrays.  According to Eq.~\eqref{nn_correlator}, throughout this phase, two-point correlation functions of the Rydberg occupation number display universal power laws at zero momentum and momentum $2{\bf Q}$, where ${\bf Q}$ is the location of the massless Dirac point.  These power-laws readily distinguish the spin liquid from competing states.  Broken-symmetry phases are generically gapped due to the absence of any continuous symmetry in the Rydberg Hamiltonian.  Power-law correlations can still arise at critical points, but contrary to the spin liquid---which is a stable phase---would entail fine-tuning.  Moreover, even at a fine-tuned critical point, emulating the precise exponents predicted by Eq.~\eqref{nn_correlator} appears unlikely.  

Descendant non-Abelian Ising topological order obtained by breaking time reversal and reflection symmetries admits a related diagnostic: The gapless chiral edge mode accompanying a gapped interior implies power-law correlations at the boundary [Eq.~\eqref{edge_expansion}] but exponentially decaying correlations in the bulk.  Following Refs.~\onlinecite{Yao_2013,Klocke2021}, one can perform more detailed diagnostics by introducing auxiliary Rydberg atoms that funnel energy into and out of the system via the chiral edge state.  This technique enables refined edge-mode characterization and, in suitable geometries, detection of individual bulk fractionalized excitations and their nontrivial braiding statistics.  

Descendant toric-code topological order is easier to generate from the parent gapless spin liquid (requiring only breaking of translation symmetry via modulation of the detuning or atom positions), yet is more challenging to detect.  Here the system is fully gapped and thus admits only exponentially decaying correlations.  At least near the gapless spin liquid phase, the dominant short-range correlations occur at zero momentum and $2{\bf Q}$.  Detecting the evolution from power-law to exponential correlations at these momenta upon ramping up translation symmetry breaking would provide one possible experimental diagnostic.  Additional supporting evidence could be obtained by deforming from a putative toric-code phase to a trivial disordered state and searching for an intervening phase transition.  

A subtle, related question concerns detection of bulk emergent fermions in the toric-code spin liquid---which is essential for readout of quantum information encoded in non-Abelian defects.  A quartet of non-Abelian defects defines a single qubit encoded in the constituent Majorana zero modes $\gamma_{1,2,3,4}$.  Defining complex fermions $f = (\gamma_1 + i\gamma_2)/2$ and $f' = (\gamma_3 + i \gamma_4)/2$, and corresponding occupation numbers $n$ and $n'$, the logical qubit states with fixed global fermion parity are $\ket{0} \equiv |n = 0 , n' = 0\rangle$ and $\ket{1} \equiv |n = 1 , n' = 1\rangle$.  Thus quantum-state readout hinges on the ability to resolve individual bulk emergent fermions.  
Developing practical measurement protocols to this end presents a very interesting future direction.

Adiabatically evolving from a trivial initial Hamiltonian to the final target presents the most straightforward way of initializing an array into either the gapless spin liquid or its descendants.  Accessing the ground state over experimentally relevant time scales could, however, prove challenging with this technique.  As a potential silver lining, we speculate that capturing finite-energy states associated with the final Hamiltonian may broaden the parameter regime in which fractionalization can be observed.  For instance, if the spin liquid is destabilized at a very small energy scale in favor of symmetry-breaking order, signatures of the former could persist at higher energies.  Semeghini et al.~invoke a similar scenario to motivate the appearnce of toric-code physics in their Rydberg array experiments \cite{RydbergSpinLiquid}.  In the solid-state context, signatures of fractionalization in Kitaev materials have been reported at intermediate energy scales even when the ground state magnetically orders \cite{TakagiQSLreview}.

Coupled-chain approaches have been previously employed to access fractionalized phases in many other settings including quantum Hall systems \cite{Kane2002,TeoKane,Mong2014,Vaezi2014,Klinovaja2014,Klinovaja2014b,Fuji2016,Kane2017,Kane2018,Fuji2019,Imamura2019,Iadecola2019,Tam2020,Tam2020b}, magnetic materials \cite{Nersesyan2003,Meng2015b,Gorohovsky2015,Patel2016,Huang2016,Huang2017,Lecheminant2017,Pereira2018,Leviatan2020}, topological insulators and superconductors \cite{Sagi2014,Meng2014,Santos2015,Mross2015,Mross2016,Sagi2017,Laubscher2019}, and more \cite{Neupert2014,Meng2015,Sagi2015,Iadecola2016,Meng2016,Halasz2017,MrossDuality,Sagi2018,Han2019,Meng2020}.  Reference~\onlinecite{Huang2017} in particular exploited Ising criticality in SU(2)-symmetric spin Hamiltonians in a manner designed to 
directly access non-Abelian Ising topological order. At least for Rydberg platforms, our method of targeting the parent gapless spin liquid (rather than the non-Abelian descendant) benefits from far simpler, symmetry-preserving microscopic inter-chain couplings.  Adapting a similar philosophy to spin systems could uncover new families of magnetic insulators realizing Kitaev-honeycomb-type spin liquid phenomenology from entirely different microscopic starting points.

\begin{acknowledgments}
It is a pleasure to thank Lesik Motrunich, David Mross, and Frederik Nathan for stimulating conversations.  We are particularly grateful to Paul Fendley for many illuminating discussions and a prior collaboration that set the foundations of our study.
The U.S. Department of Energy, Office of Science, National Quantum Information Science Research Centers, Quantum Science Center supported the construction and analysis of 2D Rydberg array models.  The Army Research Office under Grant Award W911NF-17-1-0323 supported the analysis of non-Abelian defects.  Additional support was provided by 
	the National Science Foundation through grant DMR-1848336 (RM); 
	the Caltech Institute for Quantum Information and Matter, an NSF Physics Frontiers Center with support of the Gordon and Betty Moore Foundation through Grant GBMF1250; 
	the Walter Burke Institute for Theoretical Physics at Caltech; 
	the ESQ by a Discovery Grant;
    the Gordon and Betty Moore Foundation's EPiQS Initiative, Grant GBMF8682; and
    the AFOSR YIP (FA9550-19-1-0044).
ME acknowledges support from the NSF QLCI program through grant number OMA-2016245, the DARPA ONISQ program (grant no. W911NF2010021), and the DOE Quantum Systems Accelerator Center (contract no. 7568717).
\end{acknowledgments}

\hbadness=10000	
\bibliography{main}

\end{document}